\documentclass[global,twocolumn]{svjour}

\pdfoutput=1

\usepackage{latexsym}
\usepackage{graphicx}
\usepackage{epsfig}
\usepackage{amsmath}

\journalname{}

\begin{document}
\title{Remote state preparation of a photonic quantum state via quantum teleportation}
\author{Laura T. Knoll\inst{1} \and Christian T. Schmiegelow\inst{1} \and Miguel A. Larotonda\inst{1}}
%\author{Laura T. Knoll \and Christian T. Schmiegelow \and Miguel A. Larotonda}
\institute{CEILAP, CITEDEF-CONICET, J.B. de La Salle 4397, B1603ALO Villa Martelli, Buenos Aires, Argentina
\\\email{mlarotonda@citedef.gob.ar}}
\date{September 2013}
\maketitle
\begin{abstract}
We demonstrate an experimental realization of remote state preparation via the quantum teleportation algorithm, using an entangled photon pair in the polarization degree of freedom as the quantum resource. The input state is encoded on the path of one of the photons from the pair. The improved experimental scheme allows us to control the preparation and teleportation of a state over the entire Bloch sphere with a resolution of the degree of mixture given by the coherence length of the photon pair. Both the preparation of the input state and the implementation of the quantum gates are performed in a pair of chained displaced Sagnac interferometers, which contribute to the overall robustness of the setup. An average fidelity above 0.9 is obtained for the remote state preparation process. This scheme allows for a prepared state to be transmitted on every repetition of the experiment, thus giving an intrinsic success probability of 1.
\end{abstract}
\section{Introduction}
\label{intro}
In the quantum teleportation (QT) algorithm, the sender (Alice) transmits an {\it unknown} state to the receiver (Bob) by means of a shared quantum resource, i.e. an entangled pair of qubits \cite{bennett1993teleporting}. Any implementation of the QT protocol where the input state is completely known by Alice can be regarded as remote state preparation (RSP) \cite{bennett2001remote}. RSP is equivalent to QT in the sense that both protocols have the same goal, which is to transfer a quantum state from Alice to Bob, using at least one pair of maximally entangled qubits. The difference is that in RSP, the qubit Alice transmits to Bob is {\it known}. Under certain conditions (for some specific set of qubits), in RSP the classical communication cost is lower than that required for QT \cite{pati2000minimum,lo2000classical}. In this work we use the QT as a general remote state preparation algorithm, that allows us to prepare a quantum state with a controllable degree of mixture, and send it to a distant party.

Since the development of the quantum algorithm for teleportation \cite{bennett1993teleporting}, several physical implementations were attempted, first with photons \cite{bouwmeester1997experimental,boschi1998experimental,kim2001quantum,marcikic2003long}, later on with atoms during the last decade \cite{riebe2004deterministic,barrett2004deterministic,olmschenk2009quantum}, and even on a hybrid setup, between objects of different nature \cite{sherson2006quantum}. Quantum teleportation was also used by Rosenfeld \emph{et al.} to remotely prepare a pure atomic state \cite{rosenfeld2007remote}.

In general, efforts have been focused on achieving high fidelity on the process and on increasing the physical distance between remote stages (which are of course two of the main characteristics of any envisioned quantum teleportation setup). In RSP it is also crucial to have good control over the input state. The implementation has to be able to produce superpositions of pure states, and mixed states. Peters {\it et al.}, using an entangled pair and a partial projection measurement of the polarization on the trigger qubit, showed RSP of arbitrary qubits with a scheme that is intrinsically limited to 50\% success probability due to the impossibility of implementation of a universal NOT gate \cite{peters2005remote}. The use of positive operator-valued measures (POVM) also allows for deterministic remote preparation of arbitrary pure and mixed states, at a cost of using an entangled pair and two classical bits, as demonstrated in references \cite{wu2010deterministic,killoran2010derivation}.

 In this work we present a setup for RSP that teleports a quantum state encoded on the path (or linear momentum) degree of freedom of one of the photons of a polarization-entangled pair. The use of the path qubit allows for implementation of a controlled-NOT (C-NOT) gate, which is indeed a universal gate. As a consequence, with our setup either a pure or mixed state from the Bloch sphere volume in the path qubit can be prepared with a success probability that in principle can achieve 100\%. The quantum resource is a polarization-entangled photon pair, shared by the two remote parties (Alice and Bob). An arbitrary state is prepared on the path degree of freedom of Alice's photon. This is done by means of an interferometric setup that implements an unitary gate on the path qubit \cite{englert2001universal}. Also the degree of mixture of the prepared state can be controlled by changing the temporal delay between the interferometer arms. This temporal mismatch allows us to change almost continuously between a pure superposition and a statistical mixture of states. 

We test the setup exploring several trajectories within the surface and the volume of the Bloch sphere, and we measure the average fidelity of the QT process for pure states, performing standard quantum process tomography \cite{nielsen2010quantum}.

This paper is organized as follows: first we describe the entangled photon source, then we show the photonic implementation of the QT protocol, and finally we present the results for state preparation, process tomography and fidelity of the process.

\section{Experimental Setup}
\label{setup}
\subsection{Entangled photon source}
\label{source}
Quantum teleportation relies on a quantum entanglement resource. In the present setup, the entangled system consists of a photon pair generated by spontaneous parametric downconversion (SPDC) in a BBO nonlinear crystal arrangement, pumped by a 405 nm CW diode laser. Entanglement between photons is obtained using a pair of such crystals rotated 90 degrees from each other, pumped with a beam polarized 45 degrees with respect to both crystals \cite{kwiat1999ultrabright}. The BBO crystals are cut for type-I SPDC at 29.2 degrees from the optical axis,  so that degenerated photon pairs at 810 nm emerge from the crystal at $\pm 3$ degrees from the pump beam. The total thickness of the assembly is 0.6 mm. Brightness of the source is limited by decoherence from timing information and spatial mode phase dependence generated by the birefringence of the nonlinear crystals. To avoid these problems a series of compensating crystals were introduced on the pump beam and on the photon pair paths to mitigate the longitudinal and transversal modes mismatch, respectively \cite{rangarajan2009optimizing}. The temporal pre-compensation was accomplished with a 0.3 mm a-cut $\alpha$-BBO crystal placed on the pump beam, while the transversal phase dependence was compensated using one 0.15 mm long nonlinear BBO crystals cut at the same angle of the SPDC source on each of the paths of the twin photons. With this setup we performed a Bell-type test of entanglement measuring the CHSH inequality, and we obtained a value for the estimator $S=2.703\pm0.067$. The fidelity between the produced two-photon state and the Bell state $\left | \phi ^{-}  \right \rangle =\frac{1}{\sqrt{2}}(\left | HH \right \rangle- \left | VV  \right \rangle)$ is  $F=0.945$. 
\subsection{State preparation and teleportation}
\label{rsp}
The implemented RSP protocol requires a state preparation stage and a quantum teleportation stage, both on sender side to transmit the quantum state to a distant party (Bob). The input stage is encoded on the path quantum degree of freedom of one of the photons from the entangled pair. The preparation of an arbitrary state on this spatial qubit is obtained using an interferometer with a variable relative phase between the two paths $\Delta\varphi_1$ inside the interferometer, and an additional variable relative phase $\Delta\varphi_2$ at the output \cite{englert2001universal}.

The traditional approach is to use a Mach-Zehnder interferometer. However, such optical arrangements are prone to misalignments and phase drifts due to thermal and mechanical instabilities. In order to achieve correct performance of a path qubit, an active control of the interferometer length stability has to be used. This condition increases the complexity of the setup and introduces additional noise on the photon counts due to scattering of the light from the intense control beam on all the optics elements. Instead, we have used a displaced Sagnac interferometer to prepare the input state on the path qubit \cite{schmiegelow2011selective,nagata2007beating}. This arrangement is inherently stable and allowed us to achieve 98\% interferometer visibility in coincidence with Bob detections. The displaced Sagnac interferometer also ensures that both arms have equal lengths, so that the photon wavepackets interfere with the maximum contrast by construction. As a consequence, and due to the limited length of the wavepackets ($L_{coh}\approx 21  \mu$m ), phase plates have to be inserted by pairs, one on each arm of the interferometer (phase plates are plane-parallel BK7 optical windows). Relative phase differences are generated by tilting one of the phase plates, which are mounted on goniometers and actuated via a servomotors. These compact actuators can rotate 180 degrees with a resolution of 1 degree per step. Adjusting the offset tilt so that a full $2\pi$ phase shift can be obtained, the phase resolution for a single step rotation is on the order of $\lambda/300$.

\begin{figure}[h!]
    \centering
    \includegraphics[width=0.47\textwidth]{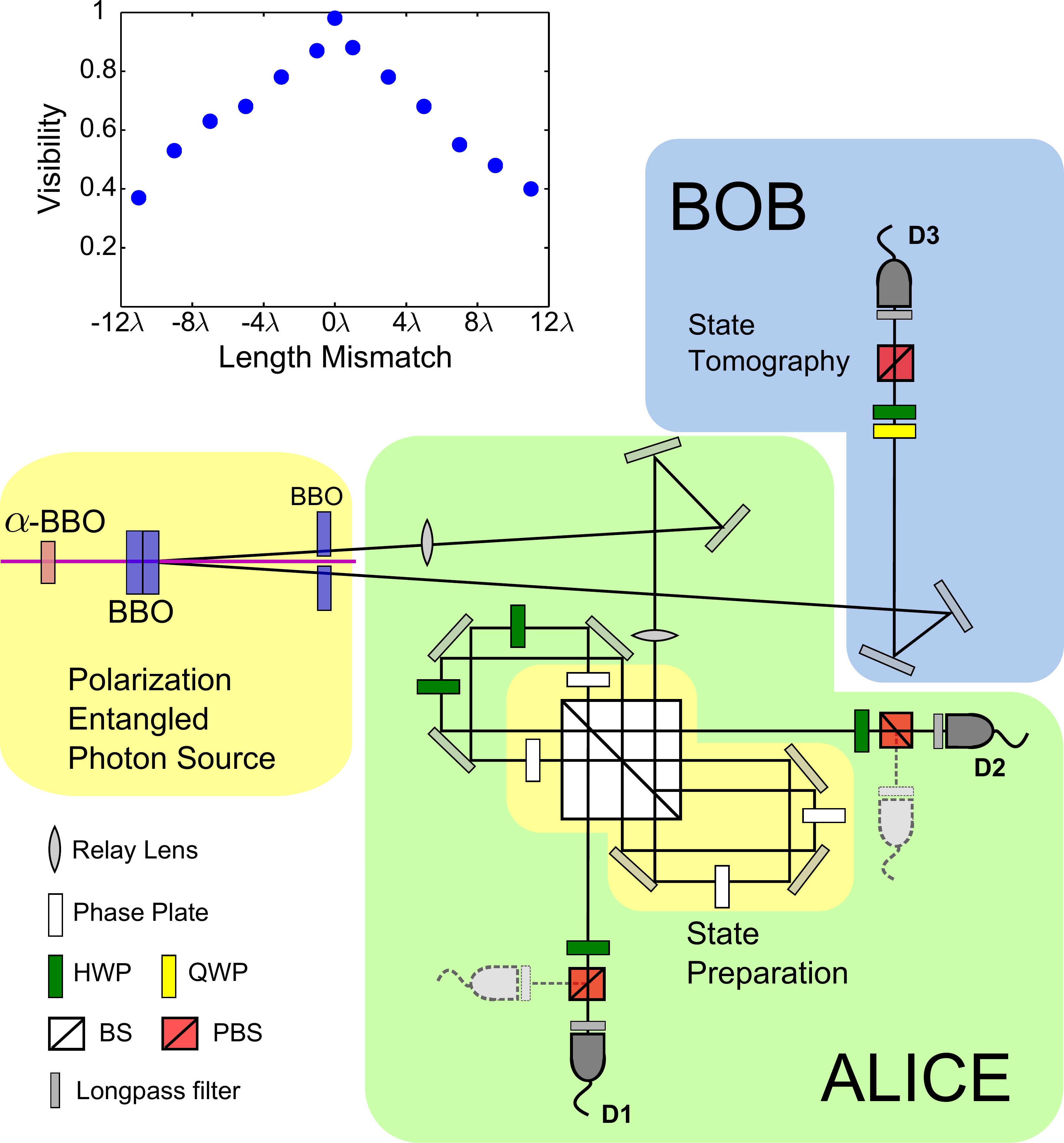}
  \caption{\footnotesize{\emph{Complete setup for remote state preparation. The SPDC entangled pair source is optimized with temporal and spatial compensating birefringent crystals. State preparation is performed on a path qubit on Alice side using a displaced Sagnac interferometer and phase plates. The C-NOT gate is implemented with half-wave plates rotated 45 degrees from each other placed on each of the logical states of the path qubit, whereas the Hadamard gate is obtained with a third passage through the beam splitter. Quantum state tomography is applied on Bob's photons in coincidence with a detection of a photon on one of Alice's outputs, to characterize the remotely prepared state. Additional detectors shown on dashed lines on Alice stage (not implemented in this work) allow for a 100\% success chance of the remote preparation process.  The inset shows the visibility curve for the first Sagnac interferometer at discrete wavelength steps, showing a FWHM width of 15 $\mu$m.}}}
  \label{fig:1}
\end{figure}

The quantum teleportation protocol is accomplished by the realization of a C-NOT gate, with the polarization qubit as target and the path qubit as the control qubit, and a Hadamard gate on the path qubit afterwards, on Alice side. A schematic representation of the experiment is depicted on figure \ref{fig:1}. The C-NOT gate is generated with two half-wave plates (HWP): one HWP with the fast axis aligned vertically on one of the paths and the other HWP with the axis rotated 45 degrees. The polarization state is left unchanged for photons on path $\left | 0  \right \rangle$ while it is rotated 90 degrees for photons on path $\left | 1  \right \rangle$. The Hadamard gate is implemented with a beam splitter. In our setup, since both paths of the prepared input qubit are emerging from a beam splitter, we again exploit the benefits of the Sagnac geometry and we fold the QT gates to form another displaced Sagnac interferometer that shares the beam splitter with the state preparation one.
 This is a compact, robust and stable setup that allows for consistent and repeatable RSP of any pure state. Furthermore, the degree of mixture of the prepared state can also be controlled: this is achieved with the inclusion of a length mismatch between the two arms of the first Sagnac interferometer. The finite coherence length of the detected photons allows us to change the interference contrast from an incoherent, non-interfering light superposition (no fringe visibility) up to maximum visibility (98\%) by tilting one of the phase plates, in steps of full-cycle phase shifts. The expanded set of accessible states with this technique is given by the density matrix
\begin{equation*}
\rho=\frac{1}{2}\begin{pmatrix}
 1+V_n\cos\left ( \Delta\varphi_1 \right ) & i e^{-i\Delta\varphi_2}V_n \sin\left ( \Delta\varphi_1 \right ) \\[2ex]
-i e^{i\Delta\varphi_2}V_n \sin\left ( \Delta\varphi_1 \right ) & 1-V_n\cos\left ( \Delta\varphi_1 \right )
\end{pmatrix},
\label{eq:instate}
\end{equation*}
where $V_n$ is the fringe visibility of the Sagnac interferometer for a phase shift of $2n\pi$. The purity if this kind of states is given by  $Tr(\rho^2)=1/2(1+V_n^2)$, which means that $V_n$ is the modulus of the Bloch vector of the prepared state. The resolution on the degree of mixture variable is proportional to the ratio $\lambda/L_{coh}$: in our case $\lambda$=810 nm and from the measured 15 $\mu$m FWHM width of the visibility curve, a length mismatch of 17$\lambda$=14 $\mu$m is needed to obtain $V=0.1$, equivalent to a purity $Tr(\rho^2)=0.505$. The visibility can be therefore adjusted with a resolution of $V_{n+1}-V_n\approx0.06$.

Measurement is done in Alice's side by projecting the two path outputs of the setup ($\left | 0  \right \rangle$ and $\left | 1  \right \rangle$) onto the canonical polarization states $\left | H  \right \rangle$, $\left | V  \right \rangle$, leading to four possible outcomes for Alice: $\left | 0H  \right \rangle$, $\left | 0V  \right \rangle$, $\left | 1H  \right \rangle$ or  $\left | 1V  \right \rangle$. In the present implementation, only two detectors are used on Alice side, and the projection on the polarization states is performed by means of half-wave plates and polarization beam splitters. This leads to an intrinsic maximum 50\% chance of success for the algorithm, although it is straightforward to increase it to a 100\% success rate with the inclusion of two additional detectors, that is, one for each of the above 2-qubit states. The remotely prepared (teleported) state is reconstructed at Bob's side by performing standard quantum state tomography of the polarization state in coincidence with any of the four projective measurements  obtained by Alice. Spatial filtering is performed upon detection, by collecting light from the outputs to single mode optical fibers before coupling them to single photon counting modules (Excelitas SPCM-AQ4C). Spectral mode filtering is accomplished using the method developed by Kurtsiefer {\it et al.} \cite{kurtsiefer2001high}: efficient light collection at a given wavelength range is achieved by matching the beam divergence with the SPDC phase matching divergence, and then matching these to the optical fiber numerical aperture. In this way, frequency filtering is transferred to the transverse spatial domain, and frequency selection can be also achieved through spatial filtering. A relay imaging system was mounted on Alice side for such task. As a consequence we only use long pass filters (1 mm-thick RG-715 Schott glass) in front of the detectors to eliminate background and pump light, and from the measured coherence length we can estimate the spectral bandwidth of the collected photon pairs to be $\Delta\lambda\approx 10$ nm. With the above described entangled pair source and a pump power of 35 mW,  a single spatial mode coincidence rate of 600 s$^{-1}$ could be obtained at the output of the complete RSP setup.

Measurement of the Stokes parameters on Bob's polarization state, needed for a tomographic reconstruction of the teleported state, is performed rotating a HWP and a quarter-wave plate in front of a beam splitter polarizer, placed before the collection optics. These wave plates are remotely actuated by servomotors which are directly coupled to the rotation axis of the wave plates, so that the angular resolution for a single-step rotation is 1 degree. All servomotors are controlled by Arduino micro controllers \cite{arduino} and operated from a desktop personal computer. 

 \section{Results}
\label{res}
\begin{figure}[h!]
    \centering
    \includegraphics[width=0.5\textwidth]{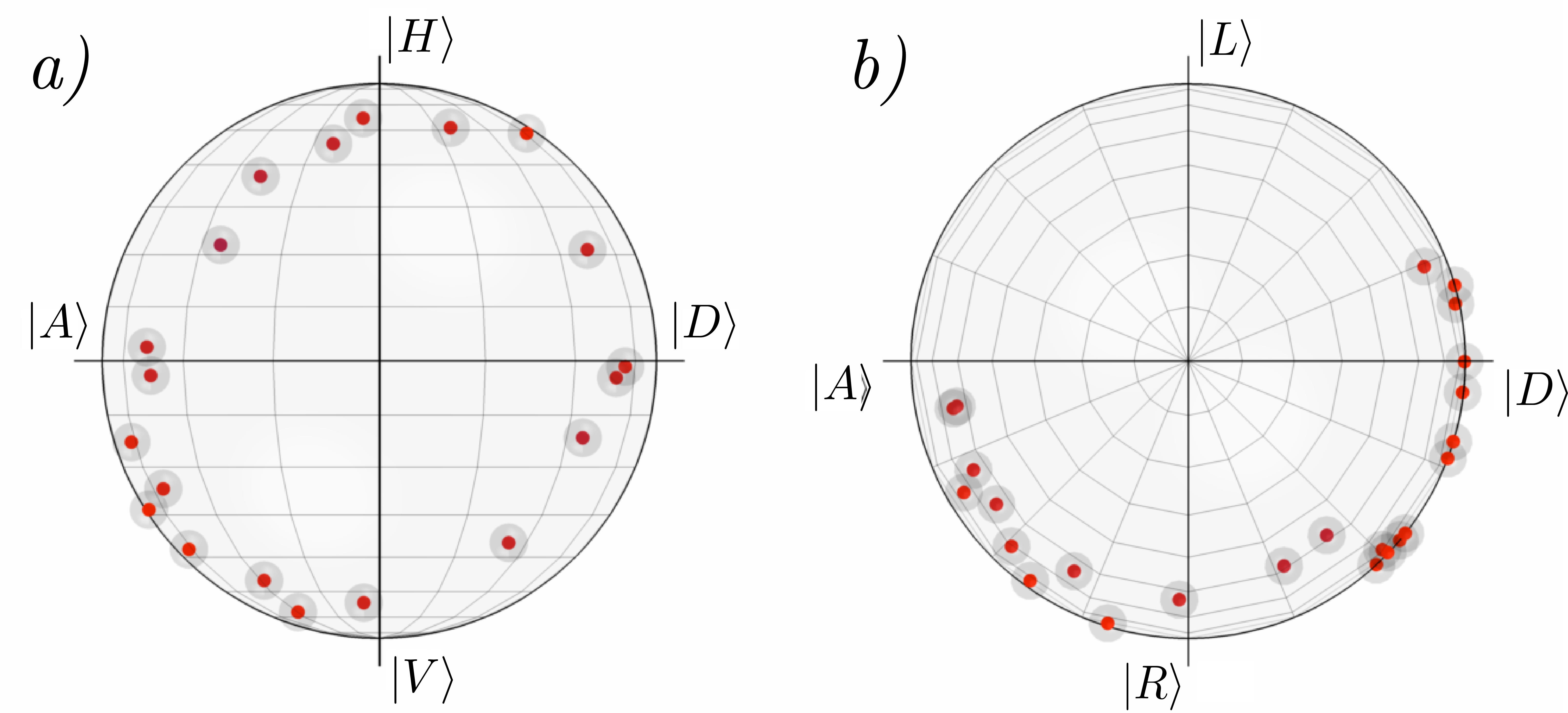}
  \caption{\footnotesize{\emph{Remote preparation of pure states along two different paths over the Bloch sphere. a) states on the $\left | H  \right \rangle \-- \left | D  \right \rangle$ meridian plane are obtained varying $\Delta\varphi_1$. b) states on the equatorial plane are obtained setting $\Delta\varphi_1$ for an equal weight superposition state and changing $\Delta\varphi_2$ to cycle through complex and real amplitudes. Grey volumes around the points represent the uncertainty of the measurements, estimated from the standard deviation of a set of similar measurements.}}}
  \label{fig:2}
\end{figure}

We evaluate the performance of our RSP setup by teleporting different sets of pure states (up to the experimental limit) from the Bloch sphere.  We perform quantum state tomography on Bob's polarization photon conditioned to the detection of  $\left | 0H  \right \rangle$ on Alice side. States from Fig. \ref{fig:2}a) are obtained setting $\Delta\varphi_2$ to produce a real, equal-weight superposition and changing $\Delta\varphi_1$ over a full $2\pi$ phase cycle. In this case all the prepared states lie on the plane generated by the $Z$ and $X$ Pauli operators eigenstates; the mean value of the {\it y}-component from the Bloch vectors of this set of teleported states is $\left \langle  p_y\right \rangle=-0.02$ with a standard deviation of 0.1. The elliptical shape and tilt that show up on the distribution of the teleported states on the $Z-X$ plane is due to the non-ideal entanglement and some residual rotation of the entangled pair used on the protocol, respectively.  The other trajectory [Fig. \ref{fig:2}b)] is obtained preparing states that have $\Delta\varphi_1$ fixed, such that the output from the first Sagnac is an equal-weight superposition of path states, while $\Delta\varphi_2$ cycles through real and complex superpositions of the canonical states: teleported states lie on the equator of the polarization Bloch sphere, with $\left \langle p_z \right \rangle=-0.02$ and a standard deviation of 0.08. In this particular case the rotation range of the servo actuator and the offset tilt needed to maximize the visibility prevent a full $2\pi$ excursion. This however can be circumvented using phase plates with increased thickness.

\begin{figure}[h!]
    \centering
    \includegraphics[width=0.43\textwidth]{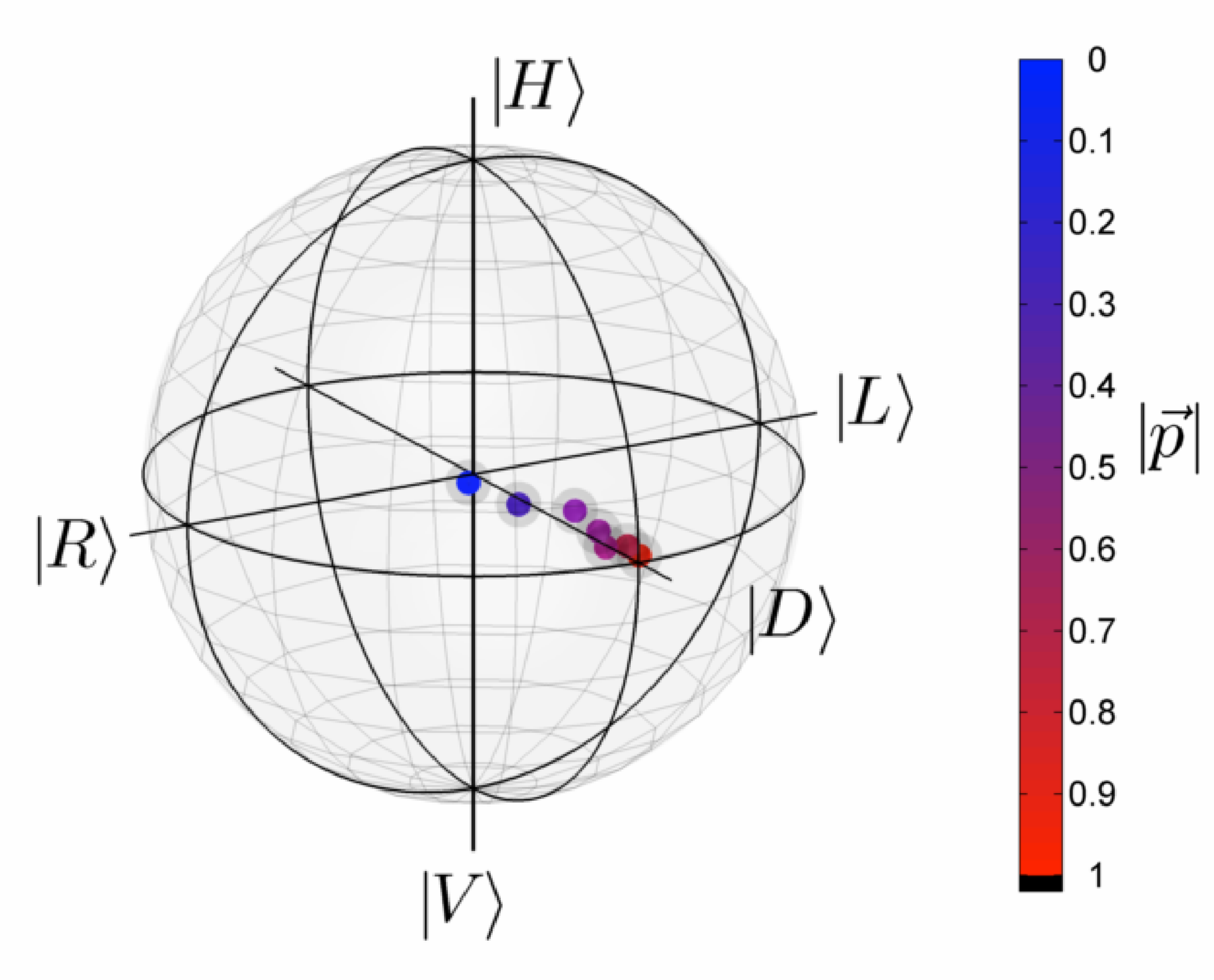}
  \caption{\footnotesize{\emph{Remote preparation of mixed states. The degree of mixture is obtained acting on the temporal mismatch between the interferometer arms. Almost pure states are obtained when the interferometer arms are compensated. Increasing the delay between arms leads to a deterioration of the fringe visibility and an increase of the degree of mixture on the prepared state. Color codes the degree of mixture of the states, given by the modulus of the Bloch vector}}}
  \label{fig:3}
\end{figure}

Preparation of mixed states is also straightforward: adding further delay between the two path states by increasing $\Delta\varphi_2$ several cycles allows us to obtain partially mixed states as the temporal overlap between wavepackets diminishes. In this way we can obtain a discrete covering of the entire Bloch sphere volume. Figure \ref{fig:3} shows teleported states between the statistical mixture $1/2\left ( \left | H  \right \rangle \left \langle H \right | + \left | V  \right \rangle \left \langle V \right | \right )$ and the coherent superposition $1/ 2 \left (\left | H  \right \rangle+\left | V  \right \rangle \right ) \left ( \left \langle H \right |+\left \langle V \right | \right ).$

The average fidelity for a single qubit quantum process described by $\mathcal E(\rho)=\sum\nolimits_{mn}\chi_{mn}E_m\rho E_n^{\dagger}$ can be calculated from the $\chi_{00}$ element of the $\chi$-matrix representation of the process \cite{renes2004symmetric}, provided that the operator base $\left \{ E_m \right \}$ satisfies: $Tr\left ( E_mE_n\right )=2\delta_{mn}$, $E_mE_m^{\dagger}=I$, and $E_0=I$ \cite{bendersky2009selective}, as

\begin{equation*}
F_{AV}=\frac{2 \chi_{00}+1}{3}.
\label{eq:avfidel}
\end{equation*}

\begin{figure}[h!]
    \centering
    \includegraphics[width=0.49\textwidth]{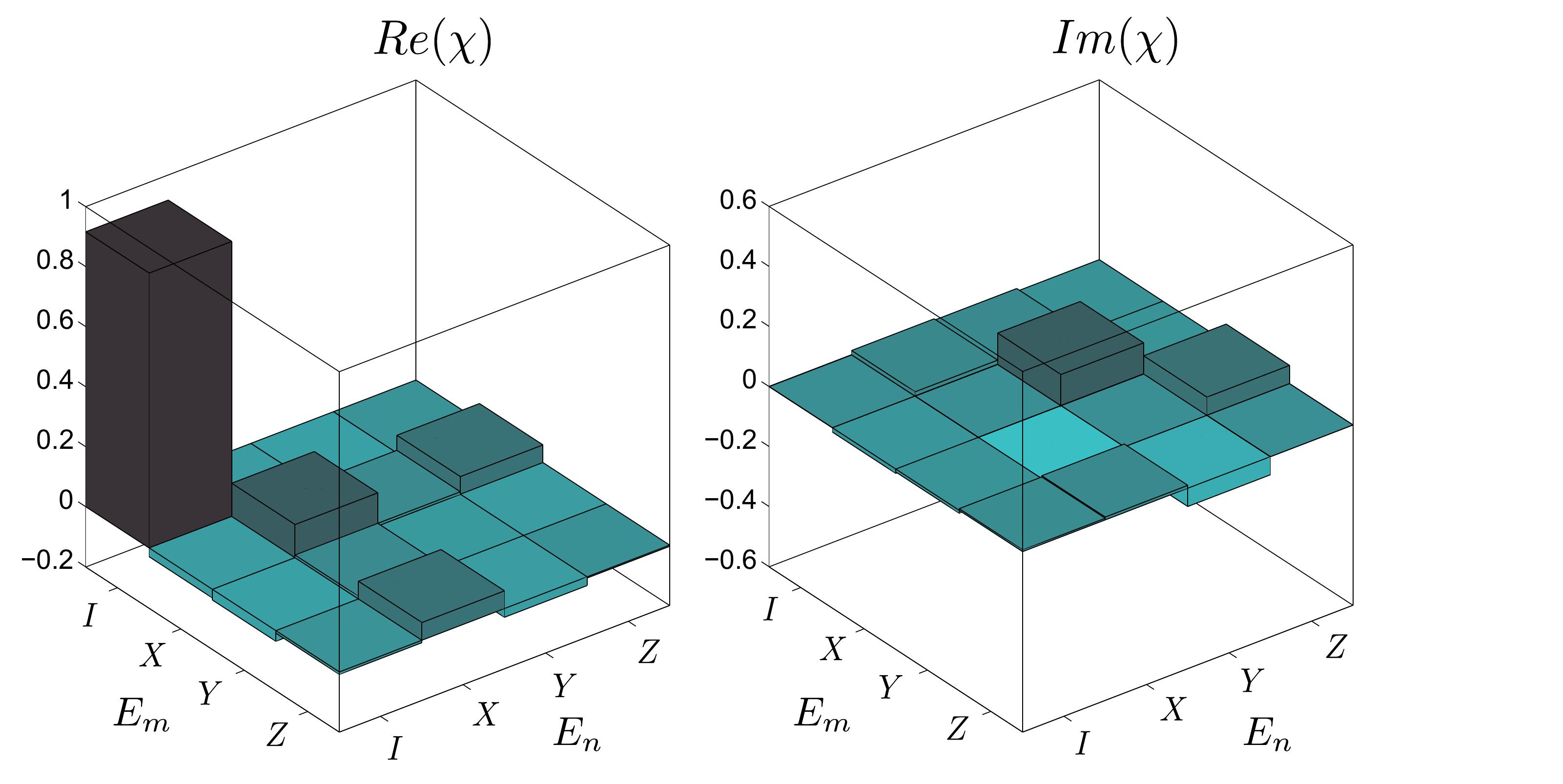}
  \caption{\footnotesize{\emph{Measured $\chi$-matrix of the teleportation process for the $\left | 0H  \right \rangle$ channel, real (left) and imaginary (right) part. This process is dominated by the identity; $\chi_{00}=0.92$, giving an average fidelity of $F_{AV}=0.95$.}}}
  \label{fig:4}
\end{figure}

Quantum process tomography was performed on each of the possible outcomes of the teleportation process, that is, the quantum state obtained by Bob conditioned to each of the four possible results obtained by Alice. Figure \ref{fig:4} plots the real and imaginary parts of the reconstructed process corresponding to detections on the $\left | 0H  \right \rangle$ channel. This output corresponds to the prepared state, meaning that the  operation required to recover the prepared state is the identity. The average fidelity for this process is  $F_{AV}^{I}=0.95$, which is well above the 2/3 limit for the maximum fidelity attainable with classical teleportation \cite{massar1995optimal}. A numerical simulation the QT protocol using the tomographically reconstructed entangled photon state produced by our source and ideally prepared input states, gives an upper limit for the average fidelity of $F_{AV}^{lim}=0.97$, which is very close to the above value. This suggests that the fidelity of the actual implemented process is mainly limited by the quality of the quantum resource. As expected, the other outputs correspond to rotations of the input state given by the Pauli operators $Z$, $Y$ and $X$. The average fidelities for these processes are  $F_{AV}^{Z}=0.87$, $F_{AV}^{Y}=0.87$ and $F_{AV}^{X}=0.90$. We ascribe these lower values for average fidelities to a state-dependent loss of contrast on the interferometer due to the residual polarization sensitivity and the unbalance of the reflection and transmission coefficients of the non-polarizing beamsplitter, in which all the four faces are used as inputs as well as outputs.

 \section{Concluding Remarks}
\label{conclu}
We report a stable and repeatable setup to remotely prepare a path qubit and transmit it to a distant polarization qubit. The experimental setup allows us to prepare both mixed and pure states. The displaced Sagnac interferometer is a key factor for the stability and control of the path qubit operations. In this work, the concept of "remote" has an additional meaning, being that all the rotations on the path qubit that are needed for the state preparation, and all the projective measurements performed on the receiver side to perform state tomography are automated via servomotors and distantly operated from a personal computer. As a consequence, mechanical and thermal disturbances are minimized. Both spectral and spatial filtering are implemented acting on the transverse profile of the photon beams. We obtain average fidelities for the remote preparation that are essentially limited by the quantum entangled resource. This implementation of RSP can be a useful tool for one-way quantum computation, which is based on a sequence of single-qubit measurements with classical feed-forward of their outcomes \cite{walther2005experimental}, and from the point of view of the teleportation algorithm, it will act as a versatile test bed for studies of fidelity enhancement or deterioration under certain noisy local environments \cite{badziag2000local,oh2002fidelity}. This work was funded by ANPCyT and CONICET grants. We thank B. Taketani and J. P. Paz for useful discussions.

%\bibliographystyle{ieeetr}
 %\bibliography{referencias}

\end{document}